\documentclass[twoside]{article}
\usepackage{moreverb}
\usepackage{fullpage}
\usepackage{fancyhdr}

\usepackage[colorlinks,bookmarksopen,bookmarksnumbered,citecolor=red,urlcolor=red]{hyperref}

\newcommand\BibTeX{{\rmfamily B\kern-.05em \textsc{i\kern-.025em b}\kern-.08em
T\kern-.1667em\lower.7ex\hbox{E}\kern-.125emX}}

\title{A Cloud Computing Survey: Developments and Future Trends in Infrastructure as a Service Computing}
\author{Jonathan Stuart Ward and Adam Barker \\School of Computer Science, University of St Andrews, UK. \\ \texttt{\{jw497, adam.barker\}st-andrews.ac.uk}}
\date{}

\begin{document}

\maketitle
\begin{abstract}
Cloud computing is a recent paradigm based around the notion of delivery of
resources via a service model over the Internet. Despite being a new 
paradigm of computation, cloud computing owes its origins to a
number of previous paradigms. The term cloud computing is well defined
and no longer merits rigorous taxonomies to furnish a definition. Instead this
survey paper considers the past, present and future of cloud computing. As an
evolution of previous paradigms, we consider the predecessors to cloud computing
and what significance they still hold to cloud services. Additionally we
examine the technologies which comprise cloud computing and how the challenges
and future developments of these technologies will influence the field.
Finally we examine the challenges that limit the growth, application and
development of cloud computing and suggest directions required to overcome
these challenges in order to further the success of cloud computing. 
\end{abstract}

\section{Introduction}
    Cloud Computing is the latest term encapsulating the delivery of computing resources as a service. It is the
current iteration of utility computing and returns to the model of `renting'
resources. The terms cloud computing and cloud are now accepted as part of industry lexicon and despite frequent misuse of these terms in advertising there is a significant body of research which underpins the area. 
Leveraging cloud computing is today, the de facto means of
deploying internet scale systems and much of the internet is tethered to a
small number of cloud providers. The advancement of cloud computing is
therefore intrinsic to the development of the next generation of internet. This
paper considers the technologies underlying cloud computing, their pasts and
their futures and the potential implications for the future of the internet. In
particular we examine the shortcomings of existing cloud systems and the
requirements of future cloud users.

    The definition of cloud computing has been well established. Where previous reviews and taxonomies have sought to provide a
clear and unambiguous definition of the domain, this is no longer necessary. The
objective of this paper is threefold. First, we examine previous analogues to cloud
computing and consider past precedent for current issues in cloud computing. Second, we examine the constituent
technologies, consider the problems within these areas and suggest the paths
for future development in cloud computing. Finally, we examine current issues and
challenges for cloud users and providers from both a technical and socio
technical perspective through specific examples.

\subsection{Scope}
``The cloud'' and ``cloud computing'' have been argued by many observers to be
ill defined and insubstantial terms. Initially dismissed by prominent
organisations including Oracle and the Free Software Foundation, Cloud
Computing has since developed into a significant and well defined domain. The most accepted
description of the general characteristics of cloud computing comes from the US
based National Institution of Standards and Technology
(NIST) and other contributers \cite{Mell}\cite{Armbrust}. It defines a concise
set of properties which define a cloud computing system:

\begin{itemize}

\item \textbf{On-demand Self Service:} A consumer is able to provision resources as needed without the need for human interaction.
\item \textbf{Broad Access:} Capabilities of a Cloud are accessed through standardised mechanisms and protocols.
\item \textbf{Resource Pooling:} The Cloud provider's resources are pooled into a shared resource which is allocated to consumers on demand.
\item \textbf{Rapid Elasticity:} Resources can be quickly provisioned and released to allow consumers to scale out and in as
required.
\item \textbf{Measured Service:} Cloud systems automatically measure a consumers use of resources allowing usage to be
monitored, controlled and reported.
\end{itemize}

    The NIST standard also defines three layers within the cloud stack:
Infrastructure as a Service (IaaS), Platform as a Service (PaaS) and Software as a Service.
Software as a service is defined as the delivery of an application, typically a
web application, on-demand via the Internet. Platform as a Service is the
delivery of a software platform and associated development tools via a service
model. Infrastructure as a Service is the provisioning of computer resources
including virtual machines (VMs), storage, networking and other resources via a
service model.  This paper refers primarily to IaaS as it is IaaS which serves as the foundation of the cloud stack and has
facilitated the phenomenon of cloud computing. While there will inevitably be
significant future development within the domains of PaaS and SaaS this
development will be highly dependent upon advances in Infrastructure as a
Service. We attempt to consider IaaS computing from both the perspective of the
consumer and the often neglected perspective of the provider.

\section{The Cousins of Cloud Computing: Similar Computing Paradigms}
\subsection{Mainframe Computing}
In many ways cloud computing has seen the industry come full circle. In the
1960s computers were extremely expensive, this prompted the development of the
mainframe computing paradigm. This paradigm saw an expensive, powerful mainframe
accessed via inexpensive terminals. The 1980s saw the rise of the PC
which largely supplanted the mainframe paradigm. Cloud computing now sees a
conceptual return to the mainframe era. In lieu of terminals, cloud computing
uses cheap consumer devices to provide cloud services. Android, iOS and
especially ChromeOS devices are inexpensive compared with regular PCs and are
designed with extensive support for cloud services. There are of course
significant differences between mainframe and cloud computing however a number
of similarities suggests value in the analogy. 

A distinct point of similarity between the mainframe era and cloud
computing is vendor lock in. IBM dominated mainframe computing and imposed
significant restrictions on the use of their software. This led to accusations
of anticompetitive practices but ultimately eliminated competition. Cloud
computing has no equivalent of IBM, public cloud services are currently dominated by
Amazon, Google, Microsoft and Rackspace. There is however little or no
interoperability between these providers' services. As a user's dependency upon
a provider's services increases it becomes increasingly difficult for them to
migrate to an alternative provider. Unlike mainframe computing, there is no
investment in hardware however the cost and difficulty of migration can be
similarly prohibitive. Increased interoperability is essential in order to
avoid the market shakeout the mainframe industry encountered in the 1970s. This is
a significant concern for the future of cloud computing.

A noteworthy point of distinction between mainframe and cloud computing is
ownership of data. In the mainframe world, ownership of data was clear.
Mainframes were owned and operated by businesses, governments and
scientific institutions. Data which resided on the mainframe was owned by the
organisation which owned the mainframe. This is not the case with cloud
computing. The devices accessing cloud services are owned by the users, the
services are owned by the providers.  No longer does the institution owning the
equipment assert ownership of the data.  This necessitates both a legal and
technical framework for asserting ownership and restricting access to cloud
hosted data. At present, ownership of data is defined only through a providers
terms of service which provides insufficient guarantee of the assumed
ownership for a considerable volume of organisations and users.

\subsection{Grid Computing}
Grid computing is conceptually similar to cloud computing and faces some of the same
challenges. While cloud computing arose from industry, grid computing traces is lineage to
academia. The objective of grid computing is to link the collective resources of multiple
independent parties to create a high performance virtual supercomputer capable of executing
computationally intensive jobs. Grid computing is typically linked to
eScience: science requiring highly distributed computation. eScience problems
typically entail substantial computation and large data sets and as such
require significant infrastructure. The bio-informatics, computational physics
and earth science communities are increasingly encountering eScience problems
and as such make heavy use of grid computing. Grids therefore, are most commonly used by the scientific community for 
performing simulations, running models, in-sillico experimentation and other complex,
eScience problems which necessitate significant resources.

The key property of grid computing which, as of yet,
is not found in cloud computing is federation. Grid computing allows for the
provisioning of a single virtual organisation (VO) from resources distributed
between a number of organisations. The VO provides dynamic resources sharing, user
authentication and access control with varying scale, structure and duration
between a number of related organisations and serves as the basis of grid
computing.

Grid computing focuses on providing high performance and high scalability. Cloud
computing alternatively focuses on delivering high scalability and low cost.
Cloud services therefore aim to provide a far lower performance to price ratio
and cannot surpass the performance of individual grid components. This is
problematic for message passing workloads which rely heavily on high
performance computers and fast interconnects. Embarrassingly Parallel workloads
however do not require high speed interconnects and scale extremely easily. The
cloud is ideal for this workload. For this reason Hadoop~\cite{Hadoop} is widely considered
as the cloud's first so called killer application. 

Gradually cloud providers are realizing the need for high performance compute
applications on the cloud. Amazon has been the first to realize this need and
offers an HPC VM instance with 10GB ethernet and substantial performance. The
lack of the preferred infiniband interconnect, slightly lowered performance and
difficulties relating to moving data to the cloud have limited adoption of HPC
clouds. Despite this, it is clear that the cloud is capable of executing
traditionally grid based workloads though not without challenge. 

A ever present bottleneck within cloud computing is the inability to scale up. This
is argued as a strength, rather cloud applications are intended to scale out. There are,
however, use cases whereby scaling up is preferable. HPC applications and other applications
well suited to grid computation often benefit from high memory and high compute servers. With
only a small number of cloud providers offering high memory and high CPU VM instances this remains
a crucial limitation. This limitation is even more significant in cloud infrastructure software
which predominantly lacks support for technologies such as Non Uniform Memory Access (NUMA) which 
allows for virtual machines to utilise the resources of several physical machines. With commodity x86 
now supporting NUMA and support for NUMA and related technologies available in the Linux
kernel since around 2005~\cite{website:linux_numa} it is now a potential area for significant research which could 
see hpc as a service become the norm, trumping even conventional cluster computing. 

The demise of grid computing in favour of cloud computing has long been
predicted.  The defining properties of grid computing: loose coupling,
heterogeneity and geographic dispersion occurred due to the need for inter
organisational cooperation. As such the grid is designed with
inter-organisational federation as a key goal. This is a property distinctly
lacking from cloud computing. While the challenges of inter-organisational
workflow management, security and governance remain unresolved by cloud computing,
grid computing will remain a significant platform for high performance
computing. 

\subsection{Cluster Computing}
Despite the prevalence of grid and cloud computing it is dedicated in-house
clusters which remain the preferred platform for HPC. The principle behind
cluster computing is simple: interconnect numerous compute nodes to provide a
high performance system. Typically this is achieved by networking large numbers
of x86 servers via a high speed Infiniband interconnect running a message
passing system to facilitate job execution. Most clusters deploy some variation
of GNU/Linux using the Message Passing Interface (MPI) or other interface. However Solaris, Mac OS X and Windows have
all been used in significant cluster deployments. 
    Clusters have a number of advantages over cloud and grid systems. Typically
clusters are owned and operated by a single authority. This allows full access
to the hardware and removes any need for federation. Full hardware access
enables users to specifically modify both the cluster and the application to achieve
optimum performance. Furthermore, the resource sharing which is crucial
in cloud computing does not take place within a cluster. An application is
executed with the full resources of the underlying hardware, not a specifically
provisioned slice. Clusters can therefore achieve significantly greater
performance than the equivalent grid or cloud solution.
    The drawbacks of cluster computing are predominantly financial. Clusters
require substantial investment and substantial maintenance, these costs are
often entirely prohibitive for smaller organisations. There exists a convention
of using a dedicated compute cluster whenever resources are available. This
convention often has numerous related groups each deploying their own
infrastructure. This can result in periods of under utilisation or idling where
a small group cannot sufficiently utilise a cluster. The resource sharing and
federation of cloud and grid computing respectively would alleviate the
problems of under utilisation by allowing for superior inter-institutional
resource usage. 

\section{The Foundations of Cloud Computing}
Cloud computing originated as a union of virtualization, distributed storage
and service oriented architecture. These three technologies have entirely
separate origins however they each encountered a renaissance in the early
2000s which led to a co-evolution. To date, major advancement within cloud
computing is attributable to advancement within one of these fields, a trend
which is set to continue. We therefore examine the origins and potential
futures and challenges of each of these technologies in an attempt to gain insight into the
future of cloud computing as a whole. 

\subsection{Virtualization}
Originating from the IBM CP/CMS operating system, virtual machines (VMs) are
one of the cornerstones of cloud computing. A VM is a software implementation
of a computer system, running in isolation alongside other processes, which
behaves as physical system.  A single multi-processor server is capable of
running several VMs, typically one per core (though cloud providers often
oversell their CPUs).  This allows for a single server
to be effectively used to capacity, reducing any unused CPU cycles and
minimising wasted energy.  Virtualizing a computer system reduces its
management overhead and allows it to be moved between physical hosts and to be
quickly instantiated or terminated.  These properties create the rapid
elasticity and scalability which underpins cloud computing. A VM is executed on
top of a hypervisor, which presents a virtual hardware platform to the VMs and
manages their execution. Historically virtualization has been a feature of
platforms with specific hardware support and remained under the purview
of mainframe computing until the late 1990s. The development of Xen in 2003 and
later the development of Intel VT-x and AMD-V, in 2005 and 2006 respectively,
made high performance x86 server virtualization feasible. This allowed for
unprecedented server consolidation and greatly decreased the time required to
provision new servers. The large scale in house deployment of virtualization at
a number of major companies is the direct catalyst for the development of cloud
computing. 

\subsection{Challenges in Virtualization}
The x86 architecture was not conceived as a platform for virtualization. The mechanisms which allow x86 based virtualization either require
a heavily modified guest OS or utilise an additional instruction set provided by modern CPUs which handles the intercepting and redirecting
traps and interrupts at the hardware level. Due to these levels of complexity there is definite performance penalty imparted through the use
of virtualization. While this penalty has considerably decreased over recent years~\cite{Menon} it still results in a virtual machine
delivering a percentage of the performance of an equivalent physical system. While some hypervisors are coming close to delivering near
native CPU performance, IO performance is still lacking. IO performance in certain scenario's suffers an 88\% slowdown compared to the
equivalent physical machine. VMs effectively trade performance for maximum utilisation of physical resources. This is non ideal for high
performance applications and is in part a motivation for the continued popularity of grid computing where non virtualized systems achieve
far greater performance. 

    Significant challenges still exist within vitalization regarding improving resource utilisation. A recent trend has been the scheduling of multiple VMs on
a single CPU core. This drastically increases the number of VMs a single host can accommodate but comes at a significant performance penalty. As 
each CPU core can execute one one VM at a time the hypervisor must switch between VMs. Each VM that is not being executed on the CPU lies idle. This
introduces IO latency as the inactive VM cannot respond to IO activity while it is inactive. Alleviating this problem is a significant research issue
as this problem significantly limits the performance of IO intensive applications, especially multimedia and real-time applications. 
    
    Improving resource utilisation is beneficial for the cloud provider and allows cheaper and greater numbers of VM instances to be made available to the consumer however it is not always without penalty. Smaller and lower
cost VM instances are also significantly problematic for many applications. In order to offer greater utilisation and lower costs many cloud providers
will schedule multiple VMs per CPU core. In the case of smaller instances there is only one CPU core available to it. In this case the host will
context switch the running VM intermittently to allow another VM to run.  Context switching a VM is a significant feat and requires the storage of
considerable state. The process of context switching imparts a significant performance overhead and has several implications for the
VMs~\cite{Barham}. This phenomena can create additional end to end delay as packets queue waiting for the recipient to return to being executed
on the CPU~\cite{Menon2005}. Furthermore it limits the ability of VMs to handle applications with real time or time sensitive applications as the
VM is not aware that for a time it is not running on the CPU and cannot account for this.  In theses cases, a less powerful non virtualized system is
better suited to the task. At present the types of application running on cloud platforms are mostly RESTful delay tolerant applications which do not
suffer significant performance or network Quality of Service (QoS) degradation given these issues. The cost and impact of context switching in virtualization is gradually
decreasing due to improved hardware support and more efficient hypervisors however the overselling of CPUs makes small cloud VM instance unsuitable
for many applications.

\subsection{Storage}
The field of databases has been dominated by SQL based relational databases for the past thirty years. SQL and relational properties provided an
appropriate model for the representation of complex information systems. The rigid structure of the relational model does not fit all problems
however. Over the past decade it has become clear that the fixed structures of tables, rows and columns are limitations when dealing with information
which is far more varied than that of traditional information systems. This had led to the development of schema-less data storage systems which lack
the conventional fixed data model. These types of systems are highly varied and typically designed for a specific use case. Despite the vast differences,
they are all united under the common identity of NoSQL databases. NoSQL, was initially not an acronym and was used to refer to database systems which 
do not employ an dialect of SQL as a query language. NoSQL has now been rechristened as "Not Only SQL" and refers to a wide array of systems~\cite{leavitt2010will}.
It is NoSQL which has been a driving force behind cloud computing. The unstructured
and highly scalable properties of many common NoSQL databases allows for large volumes of users to make use of single database installation to store
many different types of information. This is the principle behind Amazon
S3~\cite{Palankar}, Google Storage, Rackspace Files and Azure
Storage~\cite{Calder}. 

    ACID (atomicity, consistency, isolation, durability) properties are the principles which govern relational database systems and have been central to
the field since its inception. Contrary to this notion is
BASE (Basic Availability, Soft state, Eventual consistency)~\cite{Pritchett}. BASE is a notion diametrically opposed to ACID. A BASE system is
one in which requests cannot be guaranteed to be responded to, does not store data indefinitely and is not immediately consistent. ACID properties
specify properties which ensure that database transactions are processed reliably. BASE properties meanwhile specify the properties which allow for
superior performance and superior scalability. No system fully adheres to all BASE properties but rather expresses a mixture of ACID and BASE properties.
Each NoSQL system compromises at least one ACID property and therefore
expresses at least one BASE property. The exact combination of ACID and BASE
properties depends entirely upon the NoSQL solution and it's design goals.

    The CAP theorem, postulated by Brewer~\cite{Brewer} and later formally proven by Gilbert et al~\cite{Gilbert} specifies a distinct limitation for 
databases. The CAP theorem states that it is impossible for a distributed system to provide the following three guarantees:
\begin{itemize}
    \item \textbf{Consistency:} Upon a value being committed to the database the same value will always be returned unless explicitly modified.
    \item \textbf{Availability:} The database will successfully respond to all requests, regardless of failure.
    \item \textbf{Partition Tolerance:} The ability of the database to continue operating correctly in case of becoming disjointed due to network failure.
\end{itemize}

Brewer theorised that these properties are intrinsically linked and cannot be simultaneously provided. Two years later it was proven that at best a
distributed system can provide two of these three guarantees. The third property must be provided in a lesser form. This therefore entails the following taxonomy:
\begin{itemize}
    \item \textbf{Consistent and Partition Tolerant (CP):} Provides consistent data and continue to correctly operate while
        partitioned. This is achieved at a loss of the guarantee of availability. Within such systems there exists the possibility that a request may
	fail due to a node failure or other form of failure. BigTable~\cite{chang2008bigtable}, HBase, MongoDB~\cite{chodorow2010mongodb} and Reddis are all CP systems. 
    \item \textbf{Available and Partition Tolerant (AP):} Continues to service requests in the event of failure and partitioning, this is done at the cost
        of consistency. Usually this is achieved through some form of replication scheme which entails out of date replicas. These replicas are
	rendered consistent after a given period of time of inactivity. This is generally referred to as eventual consistency. Cassandra~\cite{lakshman2010cassandra}, Amazon
	Dynamo~\cite{vogels2009eventually}, Voldemort~\cite{sumbaly2012serving} and CouchDB~\cite{anderson2010couchdb} all follow this model. 
   
    \item \textbf{Consistent and Available Systems (CA):} Provides consistency and will correctly service requests if there exists no partitioning. Most
        RDBMS systems fall into this category, including MySQL and Postgres.
\end{itemize}

    Relational Database Management Systems (RDBMS) have long been the standard means of managing large volumes of structured data. The ACID and
relational properties associated with RDBMS systems are a limiting factors for many use cases. Occupying the CA portion of the Brewer taxonomy
RDBMS systems are unable to provide the same scalability as CP and AP systems.  Owing to these characteristics, RDBMs suffer from limitations in
scale, performance and fault tolerance which present a bottleneck in cloud systems~\cite{Agrawal}~\cite{Malkowski}~\cite{Wada}. In order to
achieve vast horizontal scalability and superior performance, the recent trend of NoSQL databases violate these conventions~\cite{Pokorny}. As a
result, NoSQL databases almost entirely lack a conventional relational model and most notably lack the ability to perform joins. In return for
this sacrifice NoSQL databases achieve unrivaled scalability.  Unlike traditional RDBMS which were initially conceived to operate on a single
powerful server and are not easily distributed, NoSQL databases are designed from the ground up to operate over large numbers of servers. This
allows NoSQL databases to scale through the addition of further servers.  Therefore, NoSQL databases are well suited to storing massive volumes of
non-relational, complex data which makes it well suited as a basis for cloud systems. The loss of relational and ACID properties however renders
NoSQL unfit for many use cases. 

\subsection{Challenges in Storage}
When properly designed and nominalized, RDBMS map well to physical storage mediums and can achieve noteworthy performance. While this performance is
eclipsed by that of NoSQL, that performance is gained at the expense of the relational model. Many types of data inherently lend themselves to being
represented relationally, especially data regarding people such as customer data or social network data. When these types of data are represented non
relationally, as in the case of using NoSQL, relations are often reconstructed out with the purview of the databases. While this regains some of the
lost functionality it does not fully counter for the loss of relations within the database. Furthermore as the underlying storage systems of cloud
computing rely heavily upon BASE properties there is little support for applications heavily reliant upon strong ACID compliance within the cloud. At
present, the Amazon Relational Database Service (RDS) is the predominant means of accessing a ACID compliant database in a cloud setting. Amazon RDS
exposes a web service to deploy and configure SQL databases running in a VM instance. The underlying database is otherwise typical. This does not
mitigate the problems of traditional SQL databases and will still suffer from scalability and performance issues when dealing with ``big data". For
applications which are heavily dependant upon SQL databases the only way to achieve scalability remains scaling vertically. Hence, an active problem
with the area of cloud data storage is the provisioning of relational databases. Most cloud database research has all but forgotten relational
databases and moved on to investigating NoSQL and the problems of big data.  While big data does pose significant challenges the demands of users tied
to relational databases are largely unresolved.. What is required is a new relational database developed specifically for the cloud able to scale
horizontally and offer some a greater degree of ACID properties than current NoSQL solutions.

\subsection{Service Oriented Architecture and Web Services}
Service Oriented Architecture (SOA) is in many ways an intermediate step between older concepts in distributed system and the current generation of cloud
computing systems. SOA is the practice of developing and providing software as a series of interoperable services. Services as designed as loosely
coupled units with minimum interaction between them, with each services
providing a single piece of functionality.  Individual services are then
coordinated through the process of orchestration to build an application that
utilises the services. 

The ideal of SOA is the clean
partitioning and constant representation of distributed resources \cite{Erl}.
This ideal is achieved by abstracting over previous technical and design
differences to present a universally accepted standard for the representation of
services and information. It is for this reason that cloud computing is highly
dependant upon the concept of services. SOA allows cloud computing to abstract
over the specifics of the resources being requested allowing for a standard
representation of cloud resources. 

SOA can be implemented using a number of standards including: DCOM, DDS,
CORBA, Java RMI and WCF. It is Web services however which have become a
crucial part of cloud computing. Web services are exposed over either using SOAP
messages and XML encoding over HTTP or as a RESTful service over HTTP. The combination of these technologies allows for a
very simple and open standard for service orientated communication. 

    Web service encountered an extraordinary growth in popularity in the early 2000s, largely supplanting many earlier technologies. During this time
many companies began exposing their services to developers as web services. Simultaneous developments in storage and virtualization led to the marriage
of these technologies resulting in cloud computing.  Though in many cases they are hidden behind user interfaces, it is web services which expose cloud services. 

Web services are a mature and well developed technology and as such have few
significant challenges to overcome. Web services are likely to retain their
position as the predominant means for accessing cloud services and will likely
retain their current form until the next iteration of the web.
\section{Issues for Future Cloud Computing}

\subsection{Bandwidth and Data Movement Costs}
Cloud services which are chiefly concerned with storing or operating over data
are limited by the bandwidth available to the end user.  Despite Internet
bandwidth in certain areas of the world achieving gigabit speeds, broadband
bandwidths in other regions can be as low as 500kbps. Mobile Internet bandwidth
also has the potential to be significant limited and depend on service
availability in a given area. Bandwidth limitations
pose a significant bottleneck for cloud computing. 
    Not all cloud services are bandwidth intensive, however those which are
require substantial bandwidth to achieve timely functionality.  The initial upload is often the most
significant. A user wishing to make use of a cloud storage service to store a
relatively conservative 100 GB could have to wait around 200 hours on a 2
megabit connection for the upload to complete. This problem is even greater in
the domain of mobile devices where phones, tablets and other devices attempt to
access cloud services through high latency 3G networks where delays even more
evident. Substantial delay is obviously
probative and will deter users from adopting cloud services where local
bandwidth constraints act as a bottleneck. 
    These issues are beyond the purview of cloud providers but are distinct and
substantial limitations to the accessibility of their services. For cloud
services to be considered viable alternatives to local data storage it is
essential for ubiquitous and fast broadband connections to be available. 

\subsection{Security and Trust}
Cloud computing introduces the possibility for the near universal outsourcing
of all computation and data storage requirements. The unprecedented delivery of
everything as a service brings with it a number of new security challenges. 
    Trust is an essential element of delivering everything as a service.
Confidentiality, integrity and availability of cloud hosted resources is given
only as a trust relationship between the client and the cloud provider. Trust
management is an approach to symbolically quantify decisions related to trust
by combining security policy, access control, cryptography, behavioral
analysis and artificial intelligence. The difficulties of trust management is a
significant obstacle limiting the growth of cloud computing. 
    The most significant issue of trust management is the acquisition of data
from which to derive decisions. The lower the volume of data, the less
effective the resulting decision. Part of the difficulty with cloud based
systems is that only a portion of the system is visible to the end user. The
rest of the system, which is operated by the cloud provider is inaccessible
to the end user and as such cannot be factored into trust management. This means
that any trust management decision is based on a partial view of the system and
as a result is more likely to be incorrect. 
    This is another problem whereby the interests of the user, in this case
their interest in security is at odds with the cloud providers desire to
obfuscate their infrastructure.

\subsection{Mitigating Privilege Based Attacks}
The cloud provider has total control over all operations within it's
infrastructure, therefore the integrity of user's data and software rests entirely on
their trust in the provider. There are very few technical provisions to ensure
that this trust is not violated. A rouge system administrator with root privileges
on the VM hosts can undermine all security mechanisms and obtain access to
users' applications and data. This can be easily achieved by using
\textit{libVMI} or attaching \textit{gdb} to the VM to access the memory of a
user's VM. This can allow the rouge system administrator access to private keys, plaintext
representations of data and the ability to modify any VM state. Furthermore
with physical access to the VM host, the rouge administrator can perform a
number of side channel attacks and even tamper with the hardware. 
      In order to mitigate the risk of attack it is necessary to provision a
      of closed box execution environment~\cite{Menon2005}~\cite{Menon} that ensures confidential
VM execution. It is equally necessary to provide a means to securely and
accurately attest to the confidentially of the execution environment. At
present, without such mechanisms it is impossible for a user to fully trust
that their VM instances are not subject to a privileged attack. 
    To date, no such scheme has seen been fully implemented. Despite being in the
best interests of users and encouraging greater enterprise cloud adoption the
deployment of a trusted hyper vicars is arguably not in the greatest interests of a cloud
provider. The deployment of a trusted hypervisor would require the cloud provider to expose
access to each host's trusted hypervisor and restrict their access to their own
infrastructure. The development of a trusted cloud computing environment is
therefore a trade-off between the confidentiality and  security of the users
and the amount of control cloud providers exert over their infrastructure.

\subsection{Virtual Machine Interoperability}
Cloud services are extensively based on VM formats which are specific to a
given virtualization technology results in minimal interoperability. For IaaS
clouds it is virtual machine image formats and storage formats which are the
primary point of incompatibility. Format incompatibility is further compounded
by incompatibilities in authentication, billing and resource allocation
methods. Lastly, the APIs themselves, despite being based on open standards
vary highly between cloud providers and each use alternative structures and
semantics. This incompatibility makes migration of VM, storage and other
resources between cloud providers difficult and often in the case of large
migrations, entirely unfeasible.  Significant effort has been made in
attempting to standardise aspects of IaaS cloud computing. The Open
Virtualization Format (OVF) introduced in 2007 provides a standard format for
representing VMs and is the most likely candidate for allowing VM
interoperability between IaaS providers. In addition to OVF there are standard
efforts underway by the  Distributed Management Task Force
(DMTF)~\cite{website:dmtf}, The IEEE, The Open Grid Forum~\cite{website:ogf}
and The Cloud Computing Interoperability Forum (CCIF)~\cite{parameswaran2009cloud} Which, if any, of these
standards will gain acceptance is uncertain. Each of these standards offer a
universal set of APIs and data formats for common IaaS tasks, namely the
provisioning of VMs and storage. 

Unfortunately few cloud providers offer these standardised formats. Each cloud
provider offers a number of unique features which are expressed via their own
formats and protocols and cannot be easily marshalled into a standard format.
This suggests that open standards such as OVF may never be the default formats
of IaaS clouds but rather a serialisation format to allow migration from one
service provider to another at the loss of features which cannot be represented
by the format. 

\subsection{API Interoperability} 
The largest and most influential cloud providers utilise predominantly proprietary 
and closed software. With limited collaboration and communication between providers
the earliest iterations of cloud technologies utilised entirely different protocols
and access mechanisms and were therefore largely non-interoperable~\cite{petcu2011towards}. There has been 
considerable effort invested in the development of open standards and protocol to 
facilitate API level interoperation between clouds. Amongst the initial high profile
efforts towards clouds interoperability was the Eucalyptus project~\cite{eucalyptus} which
provides an open source framework for developing private clouds which are API compatible with
Amazon web services. Eucalyptus, however, provides compatibility only with a subset of AWS features
and therefore falls short of complete interoperability. The degree of interoperability between
Eucalyptus and AWS is also noteworthy as it is now supported by an agreement between the respective
companies.

A number of other ad hoc agreements between organizations offer some degree of API compatibility between 
various, predominantly proprietary software. In each case there is less than complete API compatibility,
with obscure, legacy or new features being excluded. 
In addition to these ad hoc agreements there are a number of standards bodies which have published
sets of interoperability standards for cloud computing. These formalised standards have varying 
degrees of adoption. Organizations including the Cloud Management Unitive, the IEEE, the Cloud 
Industry Forum and the Cloud Standards Council and OASIS~\cite{parameswaran2009cloud} have either proposed or advocated the 
adoption of a set of cloud standards. Unfortunately, as is typical in the early stages of standards development there 
is a wide and often incompatible set of cloud computing standards. There are at least a dozen additional organisations either
specifically dedicated to cloud standardisation or otherwise involved in cloud standardisation which have each released
a number of draft standards~\cite{website:cloud_standards}. Few of these standards have however achieved significant 
adoption beyond niche areas. 

This vast array of potential standards has inhibited
the universal adoption of a single standard.
A likely candidate for providing a future basis for interoperability is the OpenStack project.
Openstack~\cite{website:openstack} provides a open source cloud computing platform and is backed by over 20 significant industry
bodies. In addition to providing a set of interoperability guidelines,
Openstack also implements those guidelines providing a reference implementation
for other developers. The availability of a working implementation of their own
standards has placed OpenStack in a superior position to competing standards
which have yet to have significant implementations. The availability of an
open, standardised cloud platform has seen numerous cloud providers including
Red Hat, VMWare, HP and Citrix adopt all or part of the Openstack standards within their own technologies. While the public
cloud market is still held firmly by the likes of Amazon EC2, Windows Azure and Rackspace Cloud, OpenStack
is proving to be a dominant force in the private cloud market. OpenStack has been extensively
deployed by industry, government and academia. Organisations including NASA, The US Department of Energy and HP
all operate significant private cloud deployments based on OpenStack and adhering to open standards~\cite{website:openstack:userstories} . It is therefore the case
that while other efforts continue to develop standardised APIs the best accepted standards are those of OpenStack due to the availability
of a working implementation of those standards. The viability of other standards is thus dependant upon the implementation of these standards
in real world software. Failure to provide implementations of cloud standards will inevitably see the demise of many of the current range 
of standards attempts.  

\subsection{Cloud Compliance}
Certification has long been a well accepted means to enforce compliance with a standard. Typical standards
enforce security mechanisms, performance levels and the use of specific technologies. Certification in order
to ensure compliance to a given standard is a process common to many fields. Payment processing, the storage of confidential
data and the providing services as an a affiliate of a third party organisation all frequently require some form 
of compliance process in order to obtain the necessary authorisation. Such standards are vast, complex and 
well established and many were written without cloud computing in mind. 

As such, cloud computing is incompatible with many significant standards~\cite{mather2009cloud}.  Many security security standards require 
physical access to hardware to be controlled, network communication to be isolated and all third parties barred from
accessing data. In the context of cloud computing there is no ability to manage physical access, resources are shared 
between a large pool of users and the cloud provider conceptually has access to users' data. Standards which enforce
performance requirements fare better with cloud computing but still have some limitations. In clouds where VMs are not
given exclusive access to a processor there is periodic context switching. This alone prohibits compliance to standards 
pertaining to real time applications. Furthermore the inability to guarantee exact levels of bandwidth, latency and
other metrics is prohibitive against standards requiring network guarantees.

In an initial attempt to placate users which require certain standards to be
guaranteed cloud providers provided Service Level Agreements which made
moderate claims as to security, uptime, network properties and performance. Due
to some degree of ambiguity and range of interpretations with PCI and ISO
standards some organisations which require the likes of PCI-DSS compliance have
taken SLA guarantees as adequate to maintain compliance~\cite{popovic2010cloud}. Therefore major cloud
providers have strived to achieve compliance for a number of basic standards.
Amazon Web Services, Rackspace Cloud, Azure and others have achieved certified
compliance with the PCI-DSS Level 1 and ISO 27001 security standards~\cite{subashini2011survey}.
Compliance with these standards is to perform credit card processing and the
handling of other financial data. Cloud providers' adherence to these standards
allows users who deal with such use cases to provision part of their
architecture in the cloud. These new standards will allow 'business as usual'
in the cloud but do so by removing the need for physical access, dedicated
infrastructure and other concepts which are fundamentally incompatible with
cloud computing.

There are however other, more strict standards which cloud providers have yet to achieve which prohibit other use cases
from being performed in the cloud. Data protection standards, confidentiality standards and more stringent 
financial services standards have yet to be adopted by any major cloud provider. Instead standards bodies
have begun to develop a series of standards intended specifically for cloud computing. Organisations
including the PCI, ISO, the BSI and others have begun developing and releasing new standards which avoid
inherent incompatibilities with cloud computing.

Whether cloud specific standards gain acceptance by cloud providers and whether or not relevant industries accept these
new standards as being equal to current standards will determine the success of cloud specific standards compliance.

\subsection{Government Regulation}
In 2010 following the release of a series of diplomatic cables, controversial
website Wikileaks encountered a substantial multi gigabit Distributed Denial of
Service attack. In order to mitigate the effects of this attack Wikileaks
migrated their operations to Amazon Web Services~\cite{guardian}. AWS effectively resisted the
attack and allowed Wikileaks to continue operating for several hours until
Amazon was compelled by the US government to terminate all Wikileaks operations
on AWS. This was not the first case where a government or government agency has
compelled a cloud provider to withdraw their services, it is however the
largest and most high profile incident. 
    The Wikileaks event sets an uneasy precedent. Despite Wikileaks making use
of Amazon European data center they were sanctioned under US law. One of the
often touted properties of the cloud is that data is seldom hosted in a known
location. With some services, data can at best be localised to the data center.
These creates a complex jurisdictional issue. What groups can assert control
over data and services hosted in the cloud. Case can be made for the cloud
provider, the cloud provider's government and the government of the country which
hosts the data. Without a comprehensive legal framework in place it is
impossible to conclusively argue what parties cannot access or otherwise
interfere with cloud based operations. 
    This issue is problematic for organisations such as Wikileaks which are not
well received by world governments. Unfavorable organisations can be
effectively barred from operating on the cloud by any organisations able to
exert influence against the provider. Worse still is the possibility that
governments can compel cloud providers to provide access to client's services
or data. This is a major problem for cloud computing and if this issue remains
unanswered could potentially see cloud providers relinquishing user and company data to world
governments based on a legal mandate. 

\section{Summary Of Issues and Conclusion}

The decreasing costs and increasing performance, flexibility and scalability of
cloud computing systems offers cloud providers, industry, developers and users both a
comprehensive set of advantages and a significant set of challenges:
\begin{description}
    \item[For cloud providers:] to continue delivering a cloud service requires
        significant investment in meeting the increasing demand for resources. The initial investment and total cost of ownership of cloud
        infrastructure represents a significant and increasing cost. In order to reduce these overheads and elicit future development new methods are
        required to improve resource utilisation, detect and reduce wastage and to reduce management
        complexity.
    \item[For Industry:] the lack of government and industry certification for
        cloud systems is a substantial barrier to industry cloud adoption. Outsourcing mission critical operations to a cloud provider is an
        uncomfortable paradigm for many corporations. The development of robust certification and compliance testing for cloud providers will
        alleviate some of these concerns however, further development is required to reduce the costs and complexity of managing large scale
        cloud systems. 
    
    \item[For developers:] cloud computing will allow the deployment of applications at significant scale. While at present it is possible to leverage cloud
        computing to deploy scalable applications, this is generally achieved by adapting conventional software to operate in a cloud context.
        Continued cloud development will require the abandonment of many existing programming paradigms in favour of developing applications
        specifically designed to operate at scale in the cloud.  This will also require the reevaluation of software engineering practice to provide a
        formally quantifiable approach to the design, implementation and maintenance of cloud applications.  \item[For users:] limited network access
        and limited bandwidth are significant barriers to the availability of cloud services and data. In order to ensure that cloud services are
        continually available substantial improvement to network infrastructure is required.  Furthermore the lack of common standards and robust
        security mechanisms creates the risk of vendor lock in, loss and theft of data. 
\end{description}
The late 2000s saw three separate fields co-evolve to develop cloud computing, which
has in turn become a critical and highly influential technology. Amazon EC2
alone has grown from an alternative use of Amazon's unused capacity to becoming
the largest web host in the world \cite{website:netcraft}. At present,
cloud services are used in combination with conventional services and software.
The future will see the provisioning of resources as a service become
ubiquitous. To achieve this future a number of challenges must be answered. 

The compute resources being made available on the cloud are now becoming
suitable for high performance scientific computation. However it is clear that
at present the cloud lacks the necessary federation mechanisms and sufficient
middleware platforms as to allow for the effective execution of eScience
workloads. While the middleware of grid computing can be ported to the cloud it
lacks sufficient integration with the platform and fails to offer the degree of
automation provided by the grid. The economy of cloud computing suggests that
IaaS services may be significantly cheaper than cluster or grid use for certain
workloads making cloud services a desirable option for eScience. This necessitates the development of
frameworks to provide a managed execution environment for eScience workloads on
an IaaS cloud. 

Security and confidentiality issues remain a significant challenge to
enterprise and government cloud adoption. Despite significant cloud security
research, we still lack a convincing model of trust in IaaS clouds. There is a
clear challenge remaining in developing mechanisms to provide a clear and
transparent model of security and trust for IaaS cloud services in simple and
intelligible manner. 

Cloud computing entails universal outsourcing of data. Users and businesses
have never before faced the problems of having their data stored by a third
party at such a scale. With increasing consumer and business services leveraging the cloud
model it is becoming clear that it is essential for a comprehensive legal
framework to provide an unambiguous definition as to what rights cloud
providers have to users' data. 

Conceptually cloud services afford a user superior adaptability and flexibility
compared to conventional services. Unfortunately the lack of universal
interoperability standards limits the ability of users to migrate from one
cloud service to another. There exists the significant danger of vendor lock in
when a user has committed significant resources to a cloud provider as the
costs and technical difficulties of migration may be prohibitive. To avoid the
single vendor market that existed during the mainframe era it is necessary for
cloud interoperability to be further developed and to be accepted both by users
and by cloud providers. 

Once these challenges and others have been overcome it will become feasible for
the provisioning of virtually all services and resources via a cloud computing
model. Cloud computing will eventually become the dominant platform for
Internet based hosting, storage, computation and communication and will be one
of the foundations of the next generation Internet. 
\bibliographystyle{plain}
\bibliography{ref}
\end{document}